\documentclass[12pt]{iopart}

\pdfoutput=1

\usepackage{amsmath,amsthm,amsfonts,amssymb}
\usepackage{bm,graphicx,revsymb,url}
\usepackage{colonequals}
\usepackage{cite}
\usepackage{times}

\usepackage[colorlinks={true},pdftex]{hyperref}  
\hypersetup{citecolor={blue}, filecolor={blue}, linkcolor={blue}, urlcolor={blue},
  colorlinks=true, pdfauthor={Constantin Brif, et al.},
  pdftitle={Exploring adiabatic quantum trajectories via optimal control}}

\begin{document}

\title{Exploring adiabatic quantum trajectories via optimal control}

\author{Constantin Brif\footnote[1]{Author to whom any correspondence should be addressed.}, Matthew D Grace, Mohan Sarovar and Kevin C Young}

\address{Department of Scalable \& Secure Systems Research, Sandia National Laboratories, Livermore, CA 94550, USA}

\eads{\mailto{cnbrif@sandia.gov}, \mailto{mgrace@sandia.gov}, \mailto{mnsarov@sandia.gov} and \mailto{kyoung@sandia.gov}}

\date{\today}

\begin{abstract}
Adiabatic quantum computation employs a slow change of a time-dependent control function (or functions) to interpolate between an initial and final Hamiltonian, which helps to keep the system in the instantaneous ground state. When the evolution time is finite, the degree of adiabaticity (quantified in this work as the average ground-state population during evolution) depends on the particulars of a dynamic trajectory associated with a given set of control functions. We use quantum optimal control theory with a composite objective functional to numerically search for controls that achieve the target final state with a high fidelity while simultaneously maximizing the degree of adiabaticity. Exploring properties of optimal adiabatic trajectories in model systems elucidates the dynamic mechanisms that suppress unwanted excitations from the ground state. Specifically, we discover that the use of multiple control functions makes it possible to access a rich set of dynamic trajectories, some of which attain a significantly improved performance (in terms of both fidelity and adiabaticity) through the increase of the energy gap during most of the evolution time.
\end{abstract}

\submitto{\NJP}

\maketitle


\section{Introduction}
\label{sec:Intro}

Many important, computationally challenging problems in combinatorial optimization can be solved by determining the ground state of some quantum Hamiltonian. These Hamiltonians, however, often possess spin glass structure which severely restricts the performance of thermal annealing strategies~\cite{Bapst.PhysRep.523.127.2013}. \emph{Adiabatic quantum computation} (AQC) \cite{Farhi.arXiv.quant-ph.0001106.2000, Farhi.Science.292.472.2001}, on the other hand, offers a route to producing the target ground state by slowly (adiabatically) changing the system's Hamiltonian $H(t)$ from some initial one, whose ground state is easily prepared at $t = 0$, to the final one, whose ground state encodes the solution to the problem. The quantum adiabatic theorem \cite{Born.Fock.ZPhys.51.165.1928, Kato.JPSJ.5.435.1950} guarantees that, so long as the Hamiltonian changes sufficiently slowly and the system is not subjected to external perturbations, the final state of the system will be the ground state of the problem Hamiltonian. 

By virtue of keeping the system in the instantaneous ground state of a slowly varying Hamiltonian, AQC acquires an inherent robustness to several sources of noise, such as dephasing and relaxation in the energy eigenbasis \cite{Childs.PRA.65.012322.2001}, which are known to plague the standard circuit model of quantum computation \cite{Nielsen.Chuang.2000.book}. This observation has lead to much speculation that AQC may be performed with far fewer resources than circuit model computing, for which fault tolerance demands exorbitant overheads \cite{Gottesman.arXiv.0904.2557.2009}. However, AQC's inherent robustness does not guarantee full fault tolerance, since there still exist some types of noise (in particular, single-qubit noise), which can drive undesirable transitions out of the ground state \cite{Childs.PRA.65.012322.2001, Gaitan.IJQI.4.843.2006, Amin.PRA.80.022303.2009, deVega.NJP.12.123010.2010}. Therefore, there has been much interest recently in developing error suppression and correction methods for AQC \cite{Jordan.PRA.74.052322.2006, Lidar.PRL.100.160506.2008, Paz-Silva.PRL.108.080501.2012, Young.PRX.3.041013.2013, Sarovar.NJP.15.125032.2013}. Unfortunately, a recent study found \cite{Young.PRX.3.041013.2013} that error suppression techniques based on quantum stabilizer codes \cite{Jordan.PRA.74.052322.2006, Lidar.PRL.100.160506.2008, Paz-Silva.PRL.108.080501.2012} are unlikely to be sufficient for fault-tolerant AQC. Additionally, performing effective error correction in AQC requires implementing high-weight Hamiltonian terms \cite{Young.PRX.3.041013.2013}.

A simple alternative approach to error suppression is to perform a computation faster, thus reducing the time over which errors can accumulate. However, how can faster evolution be reconciled with the requirement of adiabaticity? A variety of methods commonly known as \emph{shortcuts to adiabaticity} (see~\cite{Torrontegui.AdvAMOPhys.62.117.2013} for a recent review) aim at speeding up the transition to the target state, but at the cost of not preserving the instantaneous ground state population at intermediate times, thus forfeiting the inherent protection which is the fundamental feature of AQC. But does a decrease in evolution time necessarily have to result in a loss of population from the ground state? In this paper, we use \emph{quantum optimal control theory} (QOCT) \cite{Brif.ACP.148.1.2012, Brif.NJP.12.075008.2010, Balint-Kurti.ACP.138.43.2008, Werschnik.Gross.JPB.40.R175.2007, DAlessandro.2007.book} to explore the trade-off between the objectives of decreasing the evolution time and minimizing the population loss from the instantaneous ground state. Specifically, we consider situations where a Hamiltonian can contain multiple control functions that are allowed to vary freely in time. Then, given a finite evolution time, QOCT is used to identify control functions that maximize an objective composed of a weighted sum of two terms: one is the target-state fidelity and the other is the average ground-state population during evolution. Previous works considered different random trajectories leading to the problem Hamiltonian \cite{Farhi.arXiv.quant-ph.0208135.2002, Farhi.QIC.11.181.2011}, trajectories defined as geodesics on a manifold \cite{Mochon.PRA.75.042313.2007}, and the use of optimal control for speeding up quantum adiabatic evolution \cite{Roland.Cerf.PRA.65.042308.2002, Caneva.PRL.103.240501.2009, Caneva.PRA.84.012312.2011, Nehrkorn.arXiv.1105.1707.2011}. However, the present work is the first systematic application of QOCT to the study of adiabatic quantum trajectories.

According to the adiabatic approximation \cite{Bohm.1989.book}, the particulars of a dynamic trajectory (associated with a given set of control functions) affect the ground-state population in two ways. First, the value of the gap between the instantaneous ground state and the rest of the spectrum depends on the controls. Second, the matrix element of $\partial H/ \partial t$ depends on both the controls and their time derivatives. In general, a numerical search is required to find a control set, which implements the desired Hamiltonian interpolation in a limited time while simultaneously maximizing the degree of adiabaticity. We perform these searches, using a gradient-based optimization algorithm, for two AQC problems with a Landau--Zener type Hamiltonian. Optimization runs start from various initial control sets, including functions which are solutions to the adiabatic condition. We find that the quality of the obtained optimal control solution strongly depends on the choice of the initial set, indicating that the composite control objective has multiple local optima. The obtained results demonstrate that the richness of the dynamics accessible via the application of multiple controls makes it possible to increase the energy gap at intermediate times and thereby improve both fidelity and adiabaticity, as compared to the standard approach with a single interpolation function.

Since QOCT requires a full propagation of the system's evolution, its direct applicability is restricted to numerical studies of AQC models with a small number of qubits. However, exploring optimal AQC trajectories in model systems provides useful insights into the mechanisms that help to maintain a high degree of adiabaticity for limited evolution times. In addition, although we couch our discussion in the context of AQC, our main result --- that increasing the number of control handles in the Hamiltonian enables to preserve larger ground state populations in shorter evolution times --- is more widely applicable to all situations that employ adiabatic evolution to produce transformations of the ground state.

\section{Background}
\label{sec:back}

\subsection{Adiabatic quantum evolution}
\label{sec:AQC}

We consider a finite-dimensional closed quantum system, whose state $|\psi(t)\rangle$ satisfies the Schr\"odinger equation ($\hbar = 1$):
\begin{equation}
\label{eq:schro-psi}
\frac{\rmd}{\rmd t} |\psi(t)\rangle = -\rmi H(t) |\psi(t)\rangle ,
\end{equation}
where $H(t)$ is the time-dependent Hamiltonian of the system. It is often convenient to use the time-evolution operator $U(t) \equiv U(t,0)$, which is defined by $|\psi(t)\rangle = U(t) |\psi(0)\rangle$ and satisfies
\begin{equation}
\label{eq:schro-U}
\frac{\rmd}{\rmd t} U(t) = -\rmi H(t) U(t), \quad U(0) = I,
\end{equation}
where $I$ is the identity operator. For a system of $n$ qubits, the Hilbert space dimension is $N = 2^n$, and $U(t) \in \text{U}(N)$ (or, $U(t) \in \text{SU}(N)$ for a traceless Hamiltonian). A computation is performed by evolving the system over a finite time interval, $t \in [0,T]$. The evolution operator at the final time $T$ is denoted as $U_T \equiv U(T)$. The instantaneous eigenstates and eigenenergies of the Hamiltonian $H(t)$ are defined by 
\begin{equation}
H(t) |\phi_m(t)\rangle = E_m(t) |\phi_m(t)\rangle , \quad m = 0,1,\ldots,N-1 ,
\end{equation}
with $E_0(t) \leq E_1(t) \leq \cdots \leq E_{N-1}(t)$. 

AQC is performed by initializing the system in the ground state $|\phi_0^{(\mathrm{i})}\rangle = |\phi_0(0)\rangle$ of the initial Hamiltonian $H_{\mathrm{i}} = H(0)$ and adiabatically switching from $H_{\mathrm{i}}$ to the final Hamiltonian $H_{\mathrm{f}} = H(T)$, whose ground state $|\phi_0^{(\mathrm{f})}\rangle = |\phi_0(T)\rangle$ encodes the solution to the computational problem. In the standard formulation of AQC \cite{Farhi.arXiv.quant-ph.0001106.2000, Farhi.Science.292.472.2001}, the Hamiltonian has the form:
\begin{equation}
\label{Ham-s-interpol-1}
H(t) = (1-s) H_{\mathrm{i}} + s H_{\mathrm{f}} ,
\end{equation}
where the scaled time $s = t/T \in [0,1]$ plays the role of the interpolation parameter. It is well known \cite{Roland.Cerf.PRA.65.042308.2002, Bapst.PhysRep.523.127.2013} that the interpolation rate does not have to be constant. More generally, the Hamiltonian can be expressed as
\begin{equation}
\label{Ham-s-interpol-2}
H(t) = (1-u(t)) H_{\mathrm{i}} + u(t) H_{\mathrm{f}} ,
\end{equation}
where $u(t)$ is the interpolation function which satisfies the boundary conditions: $u(0) = 0$, $u(T) = 1$. Furthermore, controls on $H_{\mathrm{i}}$ and $H_{\mathrm{f}}$ do not need to be linearly dependent, and a more general form of the Hamiltonian is
\begin{equation}
\label{Ham-s-interpol-3}
H(t) = u_{\mathrm{i}}(t) H_{\mathrm{i}} + u_{\mathrm{f}}(t) H_{\mathrm{f}} ,
\end{equation}
where the two interpolation (control) functions satisfy the boundary conditions: $u_{\mathrm{i}}(0) = 1$, $u_{\mathrm{f}}(0) = 0$, $u_{\mathrm{i}}(T) = 0$, $u_{\mathrm{f}}(T) = 1$. These functions can be non-linearly dependent or even independent of each other. Finally, the most general form of the Hamiltonian is
\begin{equation}
\label{Ham-s-control}
H(t) = H(\{ u_k(t) \}) ,
\end{equation}
where $\{ u_k(t) \}$ ($k = 1,\ldots,K$) are time-dependent control fields (real-valued functions of time), which satisfy the boundary conditions: $H(\{ u_k(0) \}) = H_{\mathrm{i}}$ and $H(\{ u_k(T) \}) = H_{\mathrm{f}}$. In what follows, we will use the notation $u(\cdot) = \{ u_k(t) \, | \, k = 1,\ldots,K; \ t \in [0,T] \}$. A choice of the \emph{control set} $u(\cdot)$ determines the \emph{dynamic trajectory} $\psi(\cdot) = \{ |\psi(t)\rangle \, | \, t \in [0,T] \}$, and the final-time evolution operator is a functional of the controls: $U_T = U_T[u(\cdot)]$.

\subsection{Adiabatic approximation}
\label{sec:AT}

If the system is initially prepared in the ground state $|\phi_0(0)\rangle$ and the Hamiltonian varies slowly, the perturbation theory approximates the probability of the transition from $|\phi_0(0)\rangle$ to $|\phi_m (t)\rangle$ ($m \neq 0$) at time $t$ as \cite{Bohm.1989.book}
\begin{equation}
\label{eq:aa-0}
P_{0 \rightarrow m}(t) = | \langle \phi_m(t) | U(t) |\phi_0(0) \rangle |^2 \approx 
4 \frac{| \langle \phi_m(t) | \partial H/ \partial t | \phi_0 (t) \rangle |^2}{[E_m(t) - E_0(t)]^4} .
\end{equation}
Therefore, the system remains in the instantaneous ground state $|\phi_0(t)\rangle$ at all times provided that the rate of Hamiltonian change is sufficiently small, i.e.~the condition
\begin{equation}
\label{eq:aa-1}
| \langle \phi_m(t) | \partial H/ \partial t | \phi_0 (t) \rangle | \ll [E_m(t) - E_0(t)]^2 
\end{equation}
is satisfied $\forall m \neq 0$. This statement is the original formulation of the adiabatic theorem~\cite{Born.Fock.ZPhys.51.165.1928, Kato.JPSJ.5.435.1950} (more rigorous formulations of the adiabatic theorem have been given in more recent works~\cite{Nenciu.JPA.13.L15.1980, Avron.CMP.110.33.1987, Joye.Pfister.JPA.24.753.1991}; see also recent works discussing the validity of the adiabatic approximation and its bounds~\cite{Marzlin.Sanders.PRL.93.160408.2004, Tong.PRL.95.110407.2005, Ambainis.Regev.arXiv.quant-ph.0411152.2006, MacKenzie.PRA.73.042104.2006, MacKenzie.PRA.76.044102.2007, Jansen.JMP.48.102111.2007}). Under a reasonable assumption that the highest transition probability is to the first excited state, the requirement that $P_{0 \rightarrow 1}(t) \leq 4 \epsilon^2$, where $\epsilon \ll 1$ is a constant, corresponds to the \emph{adiabatic condition}:
\begin{equation}
\label{eq:aa-2}
R(t) \equiv \frac{| \langle \phi_1(t) | \partial H/ \partial t | \phi_0 (t) \rangle |}{g^2(t)} \leq \epsilon ,
\end{equation}
where $g(t) = E_1(t) - E_0(t)$ is the energy gap between the ground state and the first excited state. In the case where only one independent interpolation function is used, it is possible to solve for the unique $u(t)$ that satisfies the equality in~(\ref{eq:aa-2}) with the boundary conditions $u(0) = 0$, $u(T) = 1$. Specifically, the equality in~(\ref{eq:aa-2}) specifies a first-order differential equation for $u(t)$, whose solution depends on two parameters: the integration constant and the product $\epsilon T$; the initial and final conditions $u(0) = 0$ and $u(T) = 1$ determine the values of the integration constant and $\epsilon T$, respectively (see section~\ref{sec:problems} below for detailed examples). However, if multiple independent interpolation functions are used, there exists an infinite number of solutions that satisfy the equality in~(\ref{eq:aa-2}), since there is only one differential equation for multiple functions. For example, distinct solutions are obtained when different constraints are imposed on the functions $u_{\mathrm{i}}(t)$ and $u_{\mathrm{f}}(t)$, which eliminate one of them from~(\ref{eq:aa-2}) (e.g.~such a constraint can be a fixed form for one of the two functions or a relationship between them). We will use some of these solutions as initial control sets to start numerical searches in QOCT (see sections~\ref{sec:problems} and \ref{sec:results}).

\subsection{Quantum optimal control theory}
\label{sec:QOCT}
 
The formulation of a quantum control problem necessarily includes the definition of a quantitative control \emph{objective} (also called \emph{cost}). The objective is a functional of the controls: $J = J[u(\cdot)]$. A general class of objective functionals can be written as \cite{Brif.ACP.148.1.2012, DAlessandro.2007.book}
\begin{equation}
\label{eq:J-gen}
J[u(\cdot)] = F(U_T) + \int_0^T G(U(t),\{u_k(t)\}) \rmd t ,
\end{equation}
where $F$ is a continuously differentiable function on U($N$), and $G$ is a continuously differentiable function on $\text{U}(N) \times \mathbb{R}^K$. Usually, the first term in~(\ref{eq:J-gen}) (referred to as the \emph{final-time objective}) represents the main physical goal, while the second term (referred to as the \emph{tracking objective}) is used to incorporate various constraints on the dynamics and control fields. The optimal control problem may be stated as the search for
\begin{equation}
\label{eq:J-opt}
J^{\star} = \max_{u(\cdot)} J[u(\cdot)] ,
\end{equation}
subject to the dynamical constraint (\ref{eq:schro-U}). For the sake of consistency, we will consider only maximization of cost functionals; any control problem can be easily reformulated from minimization to maximization and vice versa by changing the sign of the functional. We will denote an optimal control set, which maximizes the objective, as $u^{\star}(\cdot)$, so that $J[u^{\star}(\cdot)] = J^{\star}$.

There are several commonly used types of the final-time objective $F(U_T)$, depending on the specific quantum control problem \cite{Brif.ACP.148.1.2012, Brif.NJP.12.075008.2010}. In particular, for state-transition control, which is relevant for AQC, the goal is to maximize the target-state fidelity defined as the probability of transition between the initial state $|\psi_{\text{i}}\rangle$ and the final (target) state $|\psi_{\text{f}}\rangle$, i.e.
\begin{equation}
\label{eq:F-Pif}
F(U_T) = P_{\text{i} \rightarrow \text{f}} = | \langle \psi_{\text{f}}| U_T |\psi_{\text{i}} \rangle |^2 .
\end{equation}

\section{Formulating optimal control theory for adiabatic quantum computation}
\label{sec:QOCT-ACQ}

In order to formulate QOCT for AQC, we need to define an objective functional. The main physical goal is to drive the system from the initial state $|\phi_0^{(\mathrm{i})}\rangle$ (the ground state of $H_{\mathrm{i}}$) into the final state $|\phi_0^{(\mathrm{f})}\rangle$ (the ground state of $H_{\mathrm{f}}$). This goal corresponds to state-transition control, with the final-time objective being the target-state fidelity of the form~(\ref{eq:F-Pif}), i.e.
\begin{equation}
\label{eq:F-aqc}
F(U_T) = | \langle \phi_0^{(\mathrm{f})} | U_T | \phi_0^{(\mathrm{i})} \rangle |^2 .
\end{equation}
Additionally, the defining feature of the AQC approach is that evolution should be sufficiently slow to keep the system in the instantaneous ground state at all times. If we employ only the final-state objective $F(U_T)$ of~(\ref{eq:F-aqc}), the adiabaticity will not be guaranteed. Therefore, the total objective functional should be of the form (\ref{eq:J-gen}), with the tracking objective (to be denoted as $J_{\text{t}}$) serving to ensure that the evolution is adiabatic. In general, there exist various possibilities for formulating the tracking objective for AQC. The approach that we focus on in this paper is to maximize the population of the instantaneous ground state averaged over the duration of evolution \cite{MacKenzie.CanJPhys.90.187.2012}, i.e.
\begin{equation}
\label{eq:Jt-a}
J_{\text{t}} = \alpha \overline{P_0} = \frac{\alpha}{T} \int_0^T  P_0(t) \rmd t .
\end{equation}
Here, 
\begin{equation}
\label{eq:IGSPop}
P_0(t) = | \langle \phi_0 (t) | \psi(t) \rangle |^2 = | \langle \phi_0 (t) | U(t) | \phi_0^{(\mathrm{i})} \rangle |^2
\end{equation}
is the instantaneous ground-state population, $\overline{P_0}$ denotes the average ground-state population, and $\alpha > 0$ is a positive weight factor that determines the relative importance of the final-time and tracking objectives (the value of $\alpha$ is selected by trial and error based on numerical optimization results). The choice of the tracking objective $J_{\text{t}}$ of the form~(\ref{eq:Jt-a}) corresponds to the goal of protecting the system against types of noise that affect excited states, including dephasing in the energy eigenbasis and leakage from the computational space. Two other possible choices of the tracking objective $J_{\text{t}}$ are discussed in~\ref{sec:alt-tos}.

Various algorithms can be used to search for an optimal control solution \cite{Brif.ACP.148.1.2012, Brif.NJP.12.075008.2010, Balint-Kurti.ACP.138.43.2008, Werschnik.Gross.JPB.40.R175.2007}. In QOCT, most numerical optimizations employ gradient-based methods \cite{Ohtsuki.JCP.120.5509.2004, Khaneja.JMR.172.296.2005, Rothman.JCP.123.134104.2005, Dominy.Rabitz.JPA.41.205305.2008, Moore.Chakrabarti.PRA.83.012326.2011, Moore.Rabitz.PRA.84.012109.2011, Palao.Kosloff.PRA.68.062308.2003, Grace.JPB.40.S103.2007, Grace.JMO.54.2339.2007, Grace.NJP.12.015001.2010, Machnes.PRA.84.022305.2011, Schirmer.Fouquieres.NJP.13.073029.2011, Reich.Ndong.Koch.JCP.136.104103.2012} (second-order methods use, in addition to the gradient, the Hessian matrix, see~\ref{sec:robust}). In order to implement such an optimization, one needs to compute the functional derivative of the objective with respect to each control field:
\begin{equation}
\label{eq:J-grad}
\frac{\delta J}{\delta u_k (t)} = \frac{\delta F}{\delta u_k (t)} + \frac{\delta J_{\text{t}}}{\delta u_k (t)} .
\end{equation}
The functional derivative of the final-time objective $F(U_T)$ of~(\ref{eq:F-aqc}) is given by (using the chain rule)
\begin{equation}
\label{eq:F-grad-1}
\frac{\delta F}{\delta u_k (t)} = 2 \text{Re} \left\{ 
\langle \phi_0^{(\mathrm{f})} | U_T | \phi_0^{(\mathrm{i})} \rangle^{\ast} 
\langle \phi_0^{(\mathrm{f})} | \frac{\delta U_T}{\delta u_k (t)} | \phi_0^{(\mathrm{i})} \rangle 
\right\} .
\end{equation}
In what follows, we assume that the Hamiltonian is linear in the controls, i.e.
\begin{equation}
\label{eq:H-linear}
H(t) = A_0 + \sum_{k=1}^{K} u_k(t) A_k ,
\end{equation}
where $A_0$ is the field-free part of the Hamiltonian and $\{ A_k \}$ are the operators through which the system couples to the control fields. Since $H(t)$ is Hermitian and $\{ u_k(t) \}$ are real, all operators $\{A_0, A_k\}$ are Hermitian. If the Hamiltonian is of the form (\ref{eq:H-linear}) and control fields are continuous functions of time, the functional derivative of the evolution operator with respect to each control field can be expressed as \cite{Ho.Rabitz.JPPA.180.226.2006, Ho.PRA.79.013422.2009}
\begin{equation}
\label{eq:Ut-grad}
\frac{\delta U(t')}{\delta u_k (t)} = \left\{ \begin{array}{ll} 
-i U(t') A_k(t), & t \leq t', \\
0, & t > t' , \end{array} \right.
\end{equation}
and, in particular, for the final-time evolution operator,
\begin{equation}
\label{eq:UT-grad}
\frac{\delta U_T}{\delta u_k (t)} = -i U_T A_k(t) , \quad 0 \leq t \leq T ,
\end{equation}
where 
\begin{equation}
A_k(t) = U^{\dagger}(t) A_k U(t)
\end{equation}
is the $k$th coupling operator in the Heisenberg picture at time $t$. In numerical simulations, time is discretized, and control fields are approximated as piecewise-constant functions of time. In such a case, results~(\ref{eq:Ut-grad}) and (\ref{eq:UT-grad}) are approximations, albeit very good ones provided that the time step is sufficiently small.\footnote{If needed, one can use a more accurate numerical method that expresses $U(t)$ as a time-ordered product of one-step propagators to compute its derivative with respect to the field value at every time step \cite{Shen.JCP.124.204106.2006}.} By substituting~(\ref{eq:UT-grad}) into~(\ref{eq:F-grad-1}), we obtain
\begin{equation}
\label{eq:F-grad-2}
\frac{\delta F}{\delta u_k (t)} = 2\, \text{Im} \left\{ 
\langle \phi_0^{(\mathrm{f})} | U_T | \phi_0^{(\mathrm{i})} \rangle^{\ast} 
\langle \phi_0^{(\mathrm{f})} | U_T A_k (t) | \phi_0^{(\mathrm{i})} \rangle 
\right\} .
\end{equation}
Using~(\ref{eq:F-grad-2}), we can compute the functional derivative of $F$ without resorting to a finite difference method.

Next, we compute the functional derivative of the tracking objective $J_{\text{t}}$ of~(\ref{eq:Jt-a}):
\begin{equation}
\label{eq:Jt-a-grad-1}
\frac{\delta J_{\text{t}} }{\delta u_k(t)} = \frac{2 \alpha}{T} \text{Re} \int_0^T 
\langle \phi_0 (t') | U(t') | \phi_0^{(\mathrm{i})} \rangle^{\ast} \frac{\delta}{\delta u_k(t)} 
\langle \phi_0 (t') | U(t') | \phi_0^{(\mathrm{i})} \rangle \rmd t' .
\end{equation}
The instantaneous ground state $|\phi_0 (t')\rangle$ depends only on field values at time $t'$, i.e.
\begin{equation}
\label{eq:phi-grad-1}
\frac{\delta |\phi_0 (t')\rangle}{\delta u_k(t)} 
= \frac{\rmd |\phi_0 (t)\rangle}{\rmd u_k(t)} \delta(t-t') .
\end{equation}
We use~(\ref{eq:phi-grad-1}) and the notation
\begin{equation}
\label{eq:phi-grad-2}
|\chi_k(t) \rangle \equiv \frac{\rmd |\phi_0 (t)\rangle}{\rmd u_k(t)} ,
\end{equation}
along with the expression~(\ref{eq:Ut-grad}) for the functional derivative of the evolution operator, to transform~(\ref{eq:Jt-a-grad-1}) into the following form:
\begin{eqnarray}
\label{eq:Jt-a-grad-2}
\frac{\delta J_{\text{t}} }{\delta u_k(t)} & = & 
\frac{2 \alpha}{T} \text{Re} \left\{ \langle \phi_0 (t) | U(t) | \phi_0^{(\mathrm{i})} \rangle^{\ast} 
\langle \chi_k (t) | U(t) | \phi_0^{(\mathrm{i})} \rangle \right\} \nonumber \\
&& + \frac{2 \alpha}{T} \text{Im} \int_t^T \langle \phi_0 (t') | U(t') | \phi_0^{(\mathrm{i})} \rangle^{\ast}
\langle \phi_0 (t') | U(t') A_k(t) | \phi_0^{(\mathrm{i})} \rangle \rmd t'  .
\end{eqnarray}
While no general expressions exist for $|\phi_0 (t)\rangle$ and $|\chi_k(t) \rangle$, they can be derived for a particular Hamiltonian model. Specifically, consider a one-qubit Hamiltonian of the form:
\begin{equation}
\label{eq:H-xz}
H(t) = x(t) \sigma_x + z(t) \sigma_z ,
\end{equation}
where $x(t)$ and $z(t)$ are the controls (real-valued functions of time), $\sigma_x$ and $\sigma_z$ are the Pauli matrices. It is straightforward to obtain
\begin{eqnarray}
&& |\phi_0 (t)\rangle = \frac{1}{\sqrt{2 h}} \begin{pmatrix} -\sqrt{h-z} \\ \sqrt{h+z} \end{pmatrix} , 
\quad 
|\phi_1 (t)\rangle = \frac{1}{\sqrt{2 h}} \begin{pmatrix} \sqrt{h+z} \\ \sqrt{h-z} \end{pmatrix} , \label{eq:phi-xz} \\ 
&& \frac{\rmd |\phi_0 (t)\rangle}{\rmd x(t)} = - \frac{z}{\sqrt{8 h^5}} \begin{pmatrix} \sqrt{h+z} \\ \sqrt{h-z} \end{pmatrix} , \quad
\frac{\rmd |\phi_0 (t)\rangle}{\rmd z(t)} = \frac{x}{\sqrt{8 h^5}} \begin{pmatrix} \sqrt{h+z} \\ \sqrt{h-z} \end{pmatrix},
\end{eqnarray}
where 
\begin{equation}
\label{eq:h-general}
h(t) = \frac{1}{2} g(t) = \sqrt{x^2(t) + z^2(t)} 
\end{equation}
is half of the energy gap $g(t)$ between the ground and excited states.

The use of the composite objective functional of the form $J = F + J_{\text{t}}$, along with a gradient-based optimization method, facilitates simple and numerically efficient searches. However, this approach has its drawbacks. It is known \cite{Chakrabarti.Wu.PRA.78.033414.2008} that QOCT methods employing such composite costs are typically incapable of identifying the genuine Pareto front that quantifies the trade-off between the individual objectives. Therefore, searches aimed at maximizing $J = F + J_{\text{t}}$ are likely to converge to solutions that underestimate the best achievable values for both the fidelity and the ground-state population. Such underperforming solutions are, in fact, local optima of $J$, which act as traps for gradient-based searches. If only the final-time objective is used (i.e.~$J = F$), then, assuming satisfaction of a few reasonable physical conditions, the control landscape $J = J[u(\cdot)]$ is free of local traps~\cite{Brif.ACP.148.1.2012}, and therefore gradient-based methods are highly effective for such types of QOCT problems \cite{Moore.Chakrabarti.PRA.83.012326.2011, Moore.Rabitz.PRA.84.012109.2011}. However, composite costs of the form $J = F + J_{\text{t}}$ often do possess local optima, due to the impossibility to simultaneously maximize both competing objectives. For optimizations involving multiple objectives, global search methods (e.g.~evolutionary strategies or genetic algorithms) can be useful \cite{Chakrabarti.Wu.PRA.78.033414.2008, Chakrabarti.Wu.PRA.77.063425.2008, Beltrani.JCP.130.164112.2009, Gollub.NJP.11.013019.2009, Sarovar.NJP.15.013030.2013, Igel.EvolComput.15.1.2007, Fonseca.EvolComput.3.1.1995, Deb.2001.book}. Still, the simplicity and efficiency of our approach make it a relevant starting point for exploring optimal AQC trajectories.

\section{AQC problems and initial control sets}
\label{sec:problems}

In order to numerically investigate the performance of the QOCT formalism presented in section~\ref{sec:QOCT-ACQ}, we consider two simple AQC problems (defined by a selection of $H_{\mathrm{i}}$ and $H_{\mathrm{f}}$), which both correspond to the one-qubit Hamiltonian~(\ref{eq:H-xz}). These two problems are presented in sections~\ref{sec:problem-I} and \ref{sec:problem-II} below.

As mentioned above, the control landscape corresponding to the composite objective $J = F + J_{\text{t}}$ is expected to possess multiple local optima, which will trap gradient-based searches starting from various initial controls. Therefore, selecting an initial control set that results in a good optimal solution (ideally, a globally optimal one) becomes a part of the optimal control problem. One approach that we explore is initializing the searches at controls that satisfy the equality in the adiabatic condition~(\ref{eq:aa-2}). For the one-qubit Hamiltonian~\eqref{eq:H-xz}, we use~\eqref{eq:phi-xz} to recast the equality in the adiabatic condition~\eqref{eq:aa-2} into the form:
\begin{equation}
\label{eq:aa-2-xz}
R(t) = \frac{| x \dot{z} - z \dot{x} |}{4 (x^2 + z^2)^{3/2}} = \epsilon .
\end{equation}
If $x(t)$ and $z(t)$ are independent functions, then, in general, equation~\eqref{eq:aa-2-xz} has an infinite number of solutions. We sample a few interesting solutions of~\eqref{eq:aa-2-xz}, obtained by selecting a constraint that eliminates one of the two functions from the equation.

\subsection{Problem~(I): $H_{\mathrm{i}} = \sigma_x$ and $H_{\mathrm{f}} = \sigma_x + \sigma_z$}
\label{sec:problem-I}

First, consider an AQC problem with $H_{\mathrm{i}} = \sigma_x$ and $H_{\mathrm{f}} = \sigma_x + \sigma_z$. The corresponding boundary conditions on the control functions are
\begin{subequations}
\label{eq:problem-I}
\begin{align}
& x(0) = 1, \ x(T) = 1 , \label{eq:problem-I-x} \\
& z(0) = 0, \ z(T) = 1 . \label{eq:problem-I-z}
\end{align}
\end{subequations}
The baseline choice for the initial control set is the linear interpolation for both functions:
\begin{equation}
\label{eq:xz-I-a}
x(s) = 1, \quad z(s) = s.
\end{equation}
The energy gap corresponding to this control set is $g(s) = 2 \sqrt{1 + s^2}$, i.e.~it monotonically increases from $g(0) = 2$ to $g(1) = 2 \sqrt{2}$.

Another possibility is to initialize the search at a control set $\{x(t),z(t)\}$, which is a solution of equation~\eqref{eq:aa-2-xz}. Since an infinite number of solutions exist, we only sample a few choices corresponding to various constraints that eliminate one of the two control functions. For problem~(I), we use constraints that fix $x(t)$ to a particular functional form, which is then substituted into~\eqref{eq:aa-2-xz} to yield a differential equation for $z(t)$. One example is the constraint that $x(t)$ is constant: $x(s) = 1, \forall s \in [0,1]$. With this constraint, \eqref{eq:aa-2-xz} reduces to
\begin{equation}
\label{eq:aa-2-xz-I-b}
\frac{\rmd z}{\rmd s} = \lambda (1 + z^2)^{3/2} ,
\end{equation}
where $\lambda = 4 \epsilon T$. By solving equation~\eqref{eq:aa-2-xz-I-b}, we obtain $z(s) = \lambda s /\sqrt{1 - \lambda^2 s^2} + C$, where $C$ is the integration constant. The boundary conditions \eqref{eq:problem-I-z} determine the values of the parameters $C$ and $\lambda$. The initial condition $z(s=0) = 0$ gives $C = 0$. The final condition $z(s=1) = 1$ takes the form $\lambda /\sqrt{1 - \lambda^2} = 1$, which gives $\lambda = 1/\sqrt{2}$ (or, equivalently, $\epsilon T = 1/\sqrt{32}$). Using these values, we obtain the control set:
\begin{equation}
\label{eq:xz-I-b} 
x(s) = 1, \quad
z(s) = \frac{s}{\sqrt{2-s^2}} , 
\end{equation}
with $R(s) = \epsilon = (\sqrt{32} T)^{-1}$. The energy gap corresponding to this control set is
\begin{equation}
\label{eq:g-I-b}
g(s) = \frac{2 \sqrt{2}}{\sqrt{2-s^2}} ,
\end{equation}
i.e.~it also monotonically increases from $g(0) = 2$ to $g(1) = 2 \sqrt{2}$.

Another possible choice is a functional form which satisfies $x(t) \geq 1$, with the purpose of increasing the gap that varies with time according to~\eqref{eq:h-general}. For example, we select: $x(s) = 1 + \sin(\pi s)$. With this constraint, \eqref{eq:aa-2-xz} reduces to
\begin{equation}
\label{eq:aa-2-xz-I-c}
\frac{\rmd z}{\rmd s} = \frac{ \pi \cos(\pi s) z 
+ \lambda \left\{[1 + \sin(\pi s)]^2 + z^2 \right\}^{3/2} }{1 + \sin(\pi s)} .
\end{equation}
By solving equation~\eqref{eq:aa-2-xz-I-c} with the boundary conditions \eqref{eq:problem-I-z}, we obtain the control set:
\begin{equation}
\label{eq:xz-I-c}
x(s) = 1 + \sin(\pi s), \quad
z(s) = \frac{[1 + \sin(\pi s)] [1 + \pi s - \cos(\pi s)]}{
\left\{ 2(\pi+2)^2 - [1 + \pi s - \cos(\pi s)]^2 \right\}^{1/2}} ,  
\end{equation}
with $R(s) = \epsilon = \pi \left[ 4 \sqrt{2} (\pi+2) T \right]^{-1}$. The energy gap corresponding to this control set is
\begin{equation}
\label{eq:g-I-c}
g(s) = \frac{2 \sqrt{2} (\pi+2) [1 + \sin(\pi s)] }{
\left\{ 2(\pi+2)^2 - [1 + \pi s - \cos(\pi s)]^2 \right\}^{1/2}} ,
\end{equation}
which reaches its maximum $g_{\max} \approx 4.34441$ at $s \approx 0.59023$.

\subsection{Problem~(II): $H_{\mathrm{i}} = \sigma_x$ and $H_{\mathrm{f}} = \sigma_z$}
\label{sec:problem-II}

Next, consider an AQC problem with $H_{\mathrm{i}} = \sigma_x$ and $H_{\mathrm{f}} = \sigma_z$. The corresponding boundary conditions on the control functions are
\begin{subequations}
\label{eq:problem-II}
\begin{align}
& x(0) = 1, \ x(T) = 0, \label{eq:problem-II-x} \\
& z(0) = 0, \ z(T) = 1 . \label{eq:problem-II-z}
\end{align}
\end{subequations}
The baseline choice for the initial control set is again the linear interpolation for both functions:
\begin{equation}
\label{eq:xz-II-a}
x(s) = 1-s, \quad z(s) = s.
\end{equation}
The energy gap corresponding to this control set is $g(s) = 2 \sqrt{1 - 2 s + 2 s^2}$, i.e.~the gap is largest at the ends of the time interval: $g(0) = g(1) = 2$, and smallest in the middle: $g(0.5) = \sqrt{2}$.

Once again, we consider the possibility to initialize the search at a control set $\{x(t),z(t)\}$, which is a solution of equation~\eqref{eq:aa-2-xz}. For problem~(II), the constraint used to eliminate one of the two independent control functions from~\eqref{eq:aa-2-xz} is a functional relationship between $x(t)$ and $z(t)$. One example is the constraint that $x(t)$ and $z(t)$ are linearly related:
\begin{equation}
\label{eq:xz-constraint-II-b}
x(s) + z(s) = 1 , \ \forall s \in [0,1] .
\end{equation}
With this constraint, \eqref{eq:aa-2-xz} reduces to
\begin{equation}
\label{eq:aa-2-xz-II-b}
\frac{\rmd z}{\rmd s} = \lambda [(1-z)^2 + z^2]^{3/2} .
\end{equation}
By solving equation~\eqref{eq:aa-2-xz-II-b} with the boundary conditions \eqref{eq:problem-II-z}, we obtain the control set:
\begin{equation}
\label{eq:xz-II-b}
x(s) = \frac{1}{2} \left[ 1 + \frac{1 - 2 s}{\sqrt{1 + 4 s - 4 s^2}} \right] , \quad
z(s) = \frac{1}{2} \left[ 1 - \frac{1 - 2 s}{\sqrt{1 + 4 s - 4 s^2}} \right] , 
\end{equation}
with $R(s) = \epsilon = 1/(2 T)$. The energy gap corresponding to this control set is
\begin{equation}
\label{eq:g-II-b}
g(s) = \frac{2}{\sqrt{1 + 4 s - 4 s^2}} ,
\end{equation}
i.e.~the gap is largest at the ends of the time interval: $g(0) = g(1) = 2$, and smallest in the middle: $g(0.5) = \sqrt{2}$. 

Another possible choice is the constraint that $x(t)$ and $z(t)$ are quadratically related: 
\begin{equation}
\label{eq:xz-constraint-II-c}
x^2(s) + z^2(s) = 1 , \ \forall s \in [0,1] .
\end{equation}
With this constraint, \eqref{eq:aa-2-xz} reduces to
\begin{equation}
\label{eq:aa-2-xz-II-c}
\frac{\rmd z}{\rmd s} = \lambda \sqrt{1 - z^2} .
\end{equation}
By solving equation~\eqref{eq:aa-2-xz-II-c} with the boundary conditions \eqref{eq:problem-II-z}, we obtain the control set:
\begin{equation}
\label{eq:xz-II-c}
x(s) = \cos(\pi s/2) , \quad 
z(s) = \sin(\pi s/2) , 
\end{equation}
with $R(s) = \epsilon = \pi/(8 T)$. The corresponding energy gap is constant: $g(s) = 2$.

\section{Numerical optimization results}
\label{sec:results}

In the numerical optimizations, we use a local, gradient-based algorithm known as D-MORPH (diffeomorphic modulation under observable-response-preserving homotopy), which was described in detail in \cite{Moore.Chakrabarti.PRA.83.012326.2011, Moore.Rabitz.PRA.84.012109.2011}. Each control field is defined on a time mesh composed of $L$ evenly spaced intervals, i.e.~$u(t) = \{ u_{\ell} | t \in (t_{\ell-1},t_{\ell}] \}_{\ell=1}^L$, where $t_{\ell} = \ell\, T/L$ (in all simulations reported here, we use $T/L = 0.01$). In this discrete representation, the control variables are the real field values $\{ u_{\ell} \}$ at the $L$ time intervals. At each step of the optimization algorithm after the initialization, each field value $u_{\ell}$ is allowed to vary freely and independently.

\begin{figure}[t]
\centering
\includegraphics[width=13.5cm]{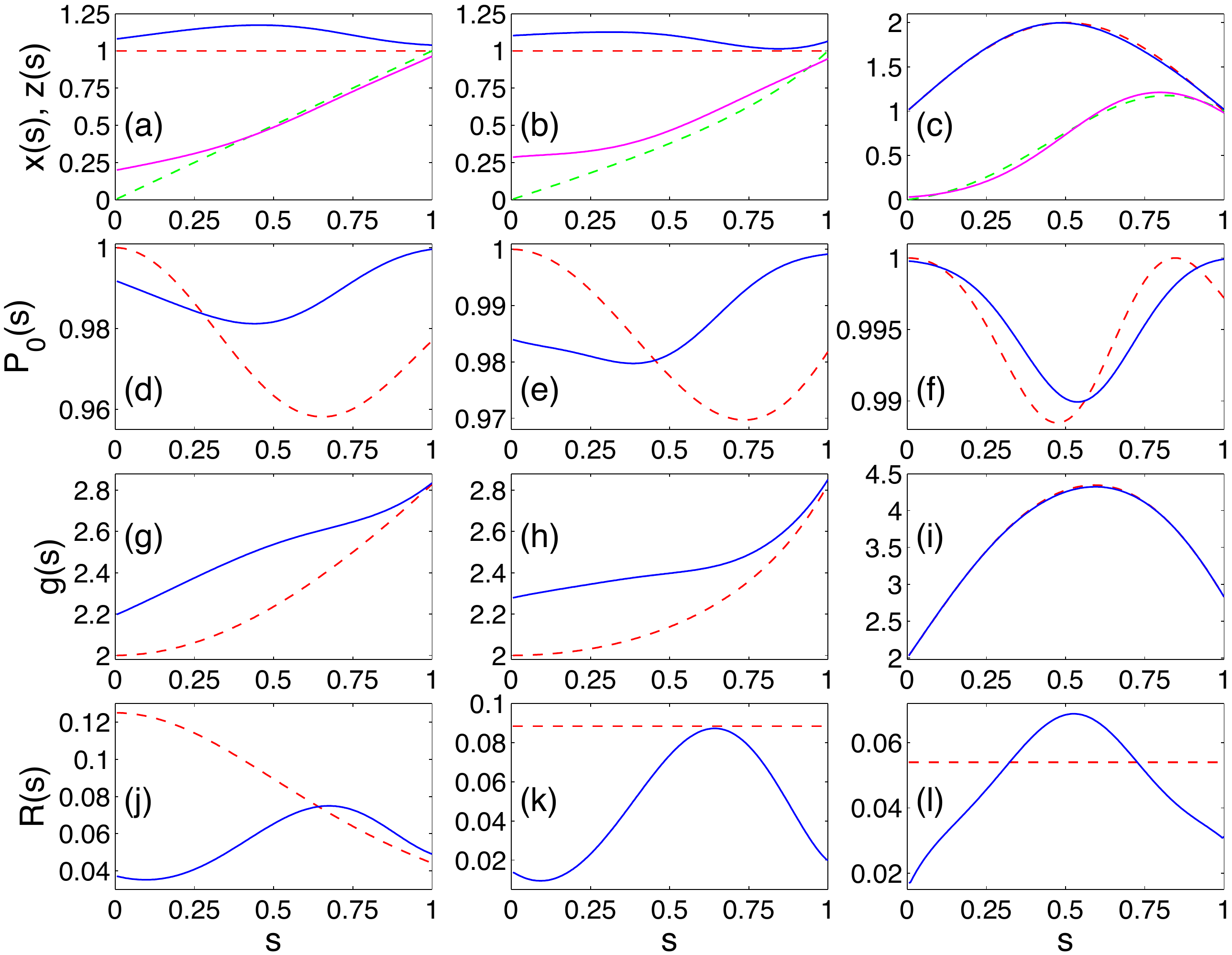}
\caption{Optimization results for the AQC problem~(I) with $H_{\mathrm{i}} = \sigma_x$ and $H_{\mathrm{f}} = \sigma_x + \sigma_z$. The control functions $x(s)$ and $z(s)$ (top row), the instantaneous ground-state population $P_0(s)$ (second row), the energy gap $g(s)$ (third row), and the adiabatic-condition ratio $R(s)$ (bottom row) are shown versus the scaled time $s = t/T$. Results are presented for optimizations starting from three initial control sets:~\eqref{eq:xz-I-a} (left column), \eqref{eq:xz-I-b} (middle column), and~\eqref{eq:xz-I-c} (right column). In all panels, red dashed lines (green dashed lines for $z(s)$ in panels (a)--(c)) show results for the initial control set and blue solid lines (magenta solid lines for $z(s)$ in panels (a)--(c)) show results for the optimal control set. In all optimizations, $T = 2$ and $\alpha = 0.1$.}
\label{fig:optim-xz-problem-I}
\end{figure}

\begin{table}[htbp]
\caption{Optimization results for the AQC problem~(I) with $H_{\mathrm{i}} = \sigma_x$ and $H_{\mathrm{f}} = \sigma_x + \sigma_z$. Three choices of the initial control set are described in the text. In all optimizations, $T = 2$.}
\label{tab:optim-xz-problem-I}
\begin{indented}
\lineup
\item[]\begin{tabular}{@{}lllllll}
\br
& \multicolumn{2}{c}{Initial set:~\eqref{eq:xz-I-a}} 
& \multicolumn{2}{c}{Initial set:~\eqref{eq:xz-I-b}}
& \multicolumn{2}{c}{Initial set:~\eqref{eq:xz-I-c}} \\
\ns
& \crule{2} & \crule{2} & \crule{2} \\ 
\multicolumn{1}{c}{$\alpha$} & 
\multicolumn{1}{c}{$1 - F$} & \multicolumn{1}{c}{$\overline{P_0}$} & 
\multicolumn{1}{c}{$1 - F$} & \multicolumn{1}{l}{$\overline{P_0}$} &  
\multicolumn{1}{c}{$1 - F$} & \multicolumn{1}{l}{$\overline{P_0}$} \\ 
\mr
\multicolumn{7}{c}{Results for initial control sets} \\
          & $2.3 \times 10^{-2}$  & 0.974 & $1.8 \times 10^{-2}$  & 0.982 & $2.8 \times 10^{-3}$  & 0.995 \\
\mr
\multicolumn{7}{c}{Results for optimal control sets} \\
$10^{0}$  & $7.1 \times 10^{-5}$  & 0.988 & $1.6 \times 10^{-5}$  & 0.987 & $1.6 \times 10^{-5}$  & 0.996 \\
$10^{-1}$ & $2.7 \times 10^{-7}$  & 0.988 & $3.9 \times 10^{-8}$  & 0.987 & $3.3 \times 10^{-7}$  & 0.996 \\
$10^{-2}$ & $4.5 \times 10^{-8}$  & 0.984 & $1.8 \times 10^{-8}$  & 0.984 & $2.9 \times 10^{-9}$  & 0.996 \\
$10^{-3}$ & $4.8 \times 10^{-10}$ & 0.984 & $1.9 \times 10^{-10}$ & 0.984 & $2.8 \times 10^{-11}$ & 0.996 \\
$10^{-4}$ & $5.7 \times 10^{-12}$ & 0.984 & $4.8 \times 10^{-12}$ & 0.984 & $1.3 \times 10^{-13}$ & 0.996 \\
$10^{-5}$ & $2.0 \times 10^{-13}$ & 0.984 & $1.1 \times 10^{-13}$ & 0.984 & $2.5 \times 10^{-14}$ & 0.996 \\
\br
\end{tabular}
\end{indented}
\end{table}

\begin{figure}[t]
\centering
\includegraphics[width=13.5cm]{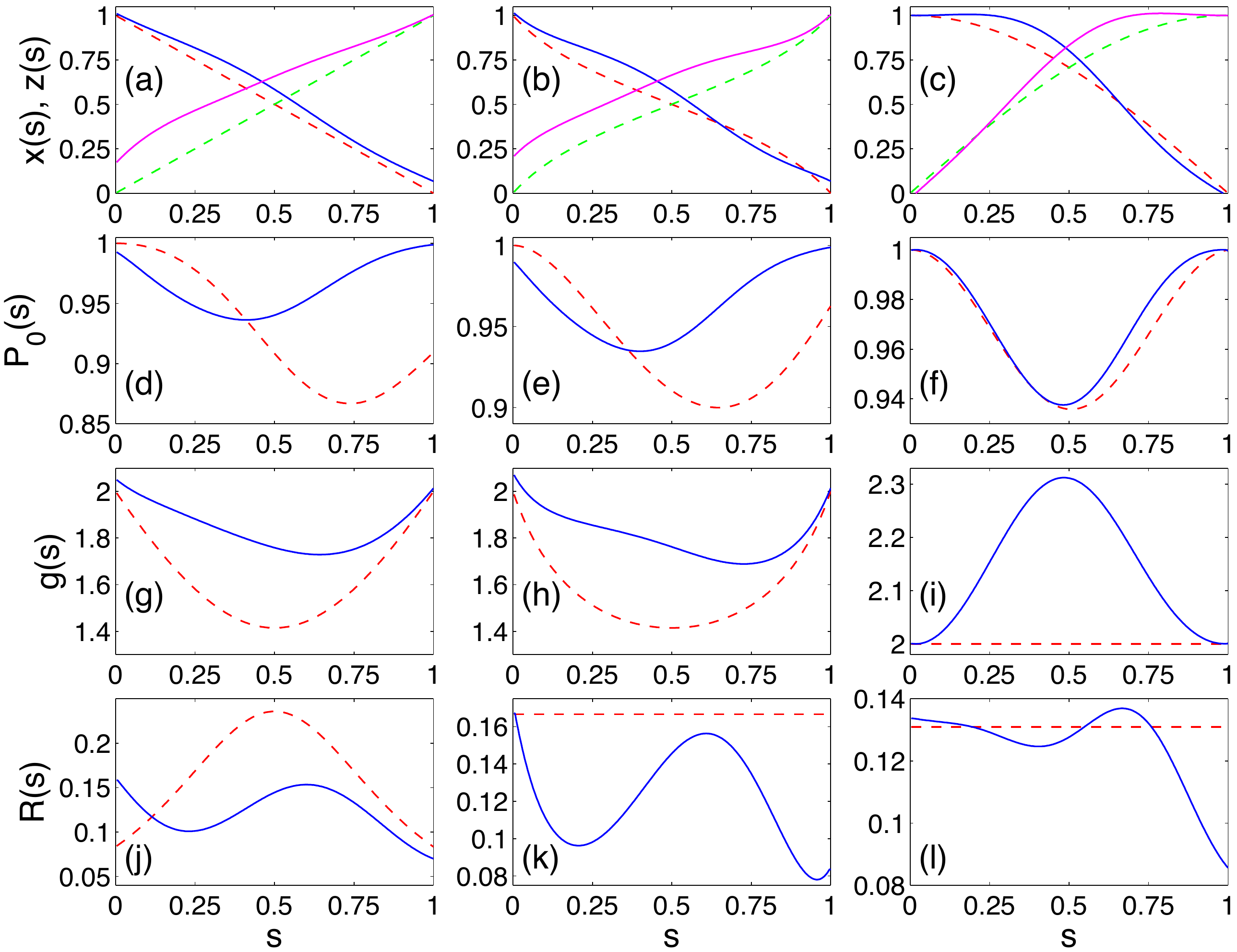}
\caption{Optimization results for the AQC problem~(II) with $H_{\mathrm{i}} = \sigma_x$ and $H_{\mathrm{f}} = \sigma_z$. The control functions $x(s)$ and $z(s)$ (top row), the instantaneous ground-state population $P_0(s)$ (second row), the energy gap $g(s)$ (third row), and the adiabatic-condition ratio $R(s)$ (bottom row) are shown versus the scaled time $s = t/T$. Results are presented for optimizations starting from three initial control sets:~\eqref{eq:xz-II-a} (left column), \eqref{eq:xz-II-b} (middle column), and~\eqref{eq:xz-II-c} (right column). In all panels, red dashed lines (green dashed lines for $z(s)$ in panels (a)--(c)) show results for the initial control set and blue solid lines (magenta solid lines for $z(s)$ in panels (a)--(c)) show results for the optimal control set. In all optimizations, $T = 3$ and $\alpha = 0.1$.}
\label{fig:optim-xz-problem-II}
\end{figure}

For problem~(I), we perform optimizations starting from the three initial control sets described in section~\ref{sec:problem-I} above:~\eqref{eq:xz-I-a},~\eqref{eq:xz-I-b},~and~\eqref{eq:xz-I-c}. For problem~(II), we perform optimizations starting from the three initial control sets described in section~\ref{sec:problem-II} above: \eqref{eq:xz-II-a}, \eqref{eq:xz-II-b}, and \eqref{eq:xz-II-c}. The search algorithm treats $x(t)$ and $z(t)$ as two independent control functions, and both are optimized.\footnote{We also examined an approach, in which only $z(t)$ is optimized, while $x(t)$ is either fixed (for problem~(I)) or computed using a functional relationship (for problem~(II)), according to the same constraint as in the definition of the respective initial control set. We found that optimizations with two independent control functions consistently attain better solutions than those with a constraint.} The optimization results are shown in figure~\ref{fig:optim-xz-problem-I} and table~\ref{tab:optim-xz-problem-I} for problem~(I) with $T = 2$, and in figure~\ref{fig:optim-xz-problem-II} and table~\ref{tab:optim-xz-problem-II} for problem~(II) with $T = 3$. The figures show the time dependence of the control functions $x(s)$ and $z(s)$, the instantaneous ground-state population $P_0(s)$, the energy gap $g(s)$, and the adiabatic-condition ratio $R(s)$, for optimizations with $\alpha = 0.1$. The tables reports values of the target-state fidelity $F$ and the average ground-state population $\overline{P_0}$, for optimizations with six various values of $\alpha$ (ranging from $1$ to $10^{-5}$).

\begin{table}[htbp]
\caption{Optimization results for the AQC problem~(II) with $H_{\mathrm{i}} = \sigma_x$ and $H_{\mathrm{f}} = \sigma_z$. Three choices of the initial control set are described in the text. In all optimizations, $T = 3$.}
\label{tab:optim-xz-problem-II}
\begin{indented}
\lineup
\item[]\begin{tabular}{@{}lllllll}
\br
& \multicolumn{2}{c}{Initial set:~\eqref{eq:xz-II-a}} 
& \multicolumn{2}{c}{Initial set:~\eqref{eq:xz-II-b}}
& \multicolumn{2}{c}{Initial set:~\eqref{eq:xz-II-c}} \\
\ns
& \crule{2} & \crule{2} & \crule{2} \\ 
\multicolumn{1}{c}{$\alpha$} & 
\multicolumn{1}{c}{$1 - F$} & \multicolumn{1}{c}{$\overline{P_0}$} & 
\multicolumn{1}{c}{$1 - F$} & \multicolumn{1}{l}{$\overline{P_0}$} &  
\multicolumn{1}{c}{$1 - F$} & \multicolumn{1}{l}{$\overline{P_0}$} \\ 
\mr
\multicolumn{7}{c}{Results for initial control sets} \\
          & $9.1 \times 10^{-2}$ & 0.925 & $3.7 \times 10^{-2}$ & 0.938 & $1.1 \times 10^{-4}$ & 0.968 \\
\mr
\multicolumn{7}{c}{Results for optimal control sets} \\
$10^{0}$  & $4.2 \times 10^{-5}$  & 0.962 & $4.6 \times 10^{-5}$  & 0.962 & $1.1 \times 10^{-4}$  & 0.968 \\
$10^{-1}$ & $4.7 \times 10^{-8}$  & 0.964 & $7.4 \times 10^{-8}$  & 0.963 & $3.8 \times 10^{-7}$  & 0.972 \\
$10^{-2}$ & $1.4 \times 10^{-8}$  & 0.955 & $2.6 \times 10^{-8}$  & 0.949 & $3.2 \times 10^{-8}$  & 0.968 \\
$10^{-3}$ & $1.5 \times 10^{-10}$ & 0.955 & $2.8 \times 10^{-10}$ & 0.948 & $3.3 \times 10^{-10}$ & 0.968 \\
$10^{-4}$ & $2.1 \times 10^{-12}$ & 0.955 & $3.4 \times 10^{-12}$ & 0.948 & $3.4 \times 10^{-12}$ & 0.968 \\
$10^{-5}$ & $1.2 \times 10^{-13}$ & 0.955 & $1.1 \times 10^{-13}$ & 0.948 & $6.9 \times 10^{-14}$ & 0.968 \\
\br
\end{tabular}
\end{indented}
\end{table}

By inspecting the results obtained for both AQC problems, we observe that the quality of an optimal control solution strongly depends on the respective initial set (i.e.~as we expected, the composite objective $J = F + J_{\text{t}}$ has multiple local optima). The performance of a dynamic trajectory, in terms of the achieved target-state fidelity and ground-state population, strongly correlates with the size of the energy gap during evolution and the rate of the Hamiltonian change in the regions where the gap is small. 

For example, consider the optimization results for problem~(I) with the initial control set~\eqref{eq:xz-I-a}, shown in the left-column panels of figure~\ref{fig:optim-xz-problem-I}. We see that, for the optimal control set, the gap abruptly increases as soon as the evolution begins (panel~(g)), resulting in a significant improvement for the ratio $R$ (panel~(j)), but this happens at the cost of a jump change of control values at the first time step (panel~(a)), leading to an immediate drop in the ground-state population (panel~(d)). A similar behavior is also observed for the optimal control set obtained by starting the search from the initial set~\eqref{eq:xz-I-b}, as shown in the middle-column panels of figure~\ref{fig:optim-xz-problem-I}. However, a very different type of dynamics is found for the initial control set~\eqref{eq:xz-I-c} and its respective optimal control set, as shown in right-column panels of figure~\ref{fig:optim-xz-problem-I}. In this case, both $x(s)$ and $z(s)$ grow from the beginning of time evolution (panel~(c)), which allows for substantially increased gap values (as compared to linear and nearly linear interpolations) at intermediate times (panel~(i)) without the need to abruptly change control values. The optimal control functions are only slightly different from the initial ones (panel~(c)), and the respective gaps are almost identical (panel~(i)), but the optimal control set utilizes a slower Hamiltonian change, as indicated by the decreased ratio $R$ (panel~(l)), in the regions near $s = 0$ and $s = 1$, i.e.~where the gap is smaller. This decrease in~$R$ translates into higher ground-state population and fidelity values (panel~(f) and table~\ref{tab:optim-xz-problem-I}).

For problem~(II), initial control sets~\eqref{eq:xz-II-a} and \eqref{eq:xz-II-b} produce a symmetric energy gap with a minimum in the middle: $g(0.5) = \sqrt{2}$. The corresponding optimization results (shown in left and middle columns, respectively, of figure~\ref{fig:optim-xz-problem-II}) indicate that the gap can be increased (panels~(g)~and~(h)), but at the cost of a jump change in the value of $z(s)$ at the first time step (panels~(a)~and~(b)) and an associated drop in the ground-state population (panels~(d) and (e)). However, once again, it is possible to find a qualitatively different dynamic regime. Specifically, the initial control set~\eqref{eq:xz-II-c} has $x(s) = \cos(\pi s/2)$ and $z(s) = \sin(\pi s/2)$, which are relatively ``flat'' near $s = 0$ and $s = 1$, respectively (panel~(c)), yielding a constant gap, $g(s) = 2$, at all times (panel~(i)). Furthermore, the optimization results reveal that by extending the regions in which $x(s)$ and $z(s)$ are ``flat'' near $s = 0$ and $s = 1$, respectively (panel~(c)), it is possible to achieve $g(s) \geq 2$ at all times, with a maximum gap value near the middle (panel~(i)). This is almost an inverse of the standard gap behavior associated with linear and nearly linear interpolations. Data shown in panels~(i)~and~(l) indicate that the improved performance of the optimal control set, as compared to the initial one, is achieved through a combination of a gap increase at intermediate times and a slower Hamiltonian change in the region near $s = 1$.

Our study demonstrates that the use of two control functions in the Hamiltonian~\eqref{eq:H-xz} makes it possible to access a broad range of dynamic trajectories. Even with only three examples of initial control sets considered here for each AQC problem, we found a significant variation of the time-dependent gap size, the adiabatic-condition ratio, and the resulting performance characteristics between different trajectories. The adiabatic condition~(\ref{eq:aa-2}) turned out to be a very useful tool that complements QOTC by providing various choices of initial control sets to seed the optimization. This strategy helped us to identify, for each AQC problem, a high-quality optimal control set whose performance is associated with a sizable increase (as compared to linear and nearly linear interpolations) of the energy gap during most of the evolution time. In each considered example, the optimization has improved the fidelity and adiabaticity, in comparison to the initial guess. While a greater relative improvement is achieved through the optimization for poorer-quality initial controls, even the best initial set benefits from the application of QOCT, especially in terms of the increased fidelity.

Results obtained for different values of the weight parameter $\alpha$ (see tables~\ref{tab:optim-xz-problem-I} and \ref{tab:optim-xz-problem-II}) reveal the competition between the final-time and tracking objectives: while fidelity values extremely close to $1$ are found when $\alpha$ is very small ($10^{-3}$ and smaller), the highest value of the average ground-state population is typically achieved for $\alpha = 0.1$. Note that a further increase of the weight parameter to $\alpha = 1$ does not improve the value of $\overline{P_0}$ for problem~(I) and even decreases it for problem~(II). This behavior is related to the fact that local optima of the composite objective $J = F + \alpha \overline{P_0}$ preclude the gradient-based searches from reaching the fidelity-adiabaticity Pareto front, and this effect may become more severe when the weight of the tracking objective is too large.

Overall, the findings of this work show that even for a simple one-qubit model there exists an extensive variety of adiabatic quantum trajectories associated with different control sets, and a proper search can discover solutions with a substantially improved performance.

\section{Conclusions and future directions}
\label{sec:concl}

The main finding of this work is that the use of multiple control functions provides access to a very rich set of dynamic trajectories, which can be explored through a strategy that combines the adiabatic condition and QOCT. This exploration helps to identify control sets that exhibit a superior performance in terms of a substantial improvement of the target-state fidelity accompanied by an increase of the average ground-state population, under a limitation on the evolution time $T$. However, the use of a composite objective of the form $J = F + J_{\text{t}}$ and a gradient-based optimization method, while simple and numerically efficient, gives only a snapshot of accessible dynamics. More insight into the attainable control performance can be gained by applying global search methods which, while being more numerically expensive, are better suited for quantifying the trade-off between multiple control objectives \cite{Chakrabarti.Wu.PRA.78.033414.2008, Chakrabarti.Wu.PRA.77.063425.2008, Beltrani.JCP.130.164112.2009, Gollub.NJP.11.013019.2009, Sarovar.NJP.15.013030.2013, Igel.EvolComput.15.1.2007, Fonseca.EvolComput.3.1.1995, Deb.2001.book}. Although such numerical studies have to be restricted to few-qubit systems, they are likely to significantly expand our understanding of the dynamic mechanisms that help to minimize the loss of adiabaticity for limited evolution times. Furthermore, global optimization methods are applicable directly in the laboratory through the use of the adaptive feedback control approach \cite{Brif.ACP.148.1.2012, Brif.NJP.12.075008.2010, Judson.Rabitz.PRL.68.1500.1992, Branderhorst.Science.320.638.2008, Biercuk.Nature.458.996.2009, Gamouras.NanoLett.13.4666.2013}.

\ack We acknowledge useful discussions with Sandia's AQUARIUS Architecture team. This work was supported by the Laboratory Directed Research and Development program at Sandia National Laboratories. Sandia is a multi-program laboratory managed and operated by Sandia Corporation, a wholly owned subsidiary of Lockheed Martin Corporation, for the United States Department of Energy's National Nuclear Security Administration under contract DE-AC04-94AL85000.

\appendix

\section{Alternative tracking objectives}
\label{sec:alt-tos}

Here, we consider a few alternative definitions of the tracking objective $J_{\text{t}}$. One additional possibility is to use the system's energy averaged over the duration of evolution as a surrogate measure for the degree of non-adiabaticity. The corresponding tracking objective is
\begin{equation}
\label{eq:Jt-b}
J_{\text{t}} = -\alpha \overline{E} = -\frac{\alpha}{T} \int_0^T E(t) \rmd t ,
\end{equation}
where
\begin{equation}
\label{eq:E-t}
E(t) = \langle \psi(t) | H(t) | \psi(t) \rangle = \langle \phi_0^{(\mathrm{i})} | U^{\dag}(t) H(t) U(t) | \phi_0^{(\mathrm{i})} \rangle 
\end{equation}
is the instantaneous energy at time $t$. For a Hamiltonian of the form~\eqref{eq:H-linear}, it is easy to find an expression for the functional derivative of the tracking objective $J_{\text{t}}$ of~\eqref{eq:Jt-b} with respect to the controls. We use~\eqref{eq:Ut-grad} and the chain rule to obtain
\begin{equation}
\label{eq:Jt-b-grad}
\frac{\delta J_{\text{t}} }{\delta u_k(t)} = 
-\frac{\alpha}{T} \langle \phi_0^{(\mathrm{i})} | A_k(t) | \phi_0^{(\mathrm{i})} \rangle 
-\frac{2 \alpha}{T} \text{Im} \int_t^T 
\langle \phi_0^{(\mathrm{i})} | U^{\dag}(t') H(t') U(t') A_k(t) | \phi_0^{(\mathrm{i})} \rangle \rmd t' .
\end{equation}
This is a nice general expression, which is straightforward to compute for any given Hamiltonian of the form \eqref{eq:H-linear}. However, our numerical studies indicate that optimal control solutions obtained with $J_{\text{t}}$ of~\eqref{eq:Jt-b} tend to underperform those obtained with $J_{\text{t}}$ of~\eqref{eq:Jt-a}.

It is also possible to attempt enforcing adiabaticity by imposing a cost on the control functions only. Specifically, adiabaticity would imply that the controls change slowly. The corresponding tracking objective should serve to minimize time derivatives of the controls, averaged over the duration of evolution and the set of all control functions:
\begin{equation}
\label{eq:Jt-c}
J_{\text{t}} = -\frac{\alpha}{K T} \sum_{k=1}^K \int_0^T  |\dot{u}_k(t)|^2 \rmd t .
\end{equation}
Unfortunately, there is no obvious way to compute the functional derivative of the tracking objective $J_{\text{t}}$ of~\eqref{eq:Jt-c} with respect to the controls. Additional numerical simulations will be needed to determine advantages and disadvantages of these and other alternative costs.

\section{Hessian of the final-time objective}
\label{sec:robust}

The Hessian matrix of the objective with respect to the controls can be used in second-order optimization methods (e.g.~the Newton--Raphson method \cite{Machnes.PRA.84.022305.2011, vonWinckel.SIAM-JSC.31.4176.2010}). Also, the Hessian plays a key role in evaluating the robustness of controlled quantum dynamics to weak random noise~\cite{Brif.QControl13.2013, Moore.PRA.86.062309.2012, Kosut.PRA.88.052326.2013}. Here, we consider the Hessian of the final-time objective $F$:
\begin{equation}
\label{eq:Hess-1}
\mathsf{H}_{k j} (t,t') = \frac{\delta^2 F}{\delta u_k(t) \delta u_j(t')} .
\end{equation}
Using the expression (\ref{eq:F-aqc}) for $F$, the Hessian can be written as
\begin{eqnarray}
\label{eq:Hess-2}
\mathsf{H}_{k j} (t,t') & = & 2\, \text{Re} \left[
\langle \phi_0^{(\mathrm{f})} | U_T | \phi_0^{(\mathrm{i})} \rangle^{\ast} 
\langle \phi_0^{(\mathrm{f})} | \frac{\delta^2 U_T}{\delta u_k(t) \delta u_j(t')} | \phi_0^{(\mathrm{i})} \rangle 
\right. \nonumber \\
&& + \left. \langle \phi_0^{(\mathrm{f})} | \frac{\delta U_T}{\delta u_k(t)} | \phi_0^{(\mathrm{i})} \rangle^{\ast} 
\langle \phi_0^{(\mathrm{f})} | \frac{\delta U_T}{ \delta u_j(t')} | \phi_0^{(\mathrm{i})} \rangle
\right] .
\end{eqnarray}
In numerical simulations, when time is discretized and control functions are treated as piecewise constant, the functional derivatives of $U_T$ in~(\ref{eq:Hess-2}) and, correspondingly, the Hessian can be evaluated via a straightforward numerical procedure \cite{Shen.JCP.124.204106.2006}. If the controls are continuous functions of time and enter linearly into the Hamiltonian, the functional derivatives can be expressed analytically. Specifically, assuming that the Hamiltonian has the form~(\ref{eq:H-linear}) and control functions are continuous, one obtains \cite{Ho.Rabitz.JPPA.180.226.2006, Ho.PRA.79.013422.2009} the expression~(\ref{eq:UT-grad}) for $\delta U_T/\delta u_k(t)$ and
\begin{equation}
\label{eq:U-der-2}
\frac{\delta^2 U_T}{\delta u_k(t) \delta u_j(t')} = - U_T A_j(t') A_k(t) .
\quad \ 0 \leq t \leq t' \leq T .
\end{equation}
The result for $t > t'$ is obtained by reversing the order of $A_k(t)$ and $A_j(t')$ in~(\ref{eq:U-der-2}). 

Next, we want to evaluate the Hessian at the optimum: $\mathsf{H}_{k j}^{\star} (t,t') = \left. \mathsf{H}_{k j}(t,t') \right|_{u(\cdot) = u^{\star}(\cdot)}$. Assume that the control set $u^{\star}(\cdot)$ is optimal not only for the total objective $J$, but also for its final-time part $F$, so that controlled evolution produces the target state:
\begin{equation}
\label{eq:U-opt}
U_T[u^{\star}(\cdot)] | \phi_0^{(\mathrm{i})} \rangle = | \phi_0^{(\mathrm{f})} \rangle .
\end{equation}
Substituting~(\ref{eq:UT-grad}) and (\ref{eq:U-der-2}) into~(\ref{eq:Hess-2}) and evaluating at the optimum with the assumption (\ref{eq:U-opt}), we obtain:
\begin{equation}
\label{eq:Hess-3}
\mathsf{H}^{\star}_{k j} (t,t') = 
- \langle \phi_0^{(\mathrm{i})} | [A_k(t) A_j(t') + A_j(t') A_k(t)] | \phi_0^{(\mathrm{i})} \rangle 
+ 2 \langle \phi_0^{(\mathrm{i})} | A_k(t) | \phi_0^{(\mathrm{i})} \rangle 
\langle \phi_0^{(\mathrm{i})} | A_j(t') | \phi_0^{(\mathrm{i})} \rangle .
\end{equation}
Note that the expression~(\ref{eq:Hess-3}) is symmetric in $A_k(t)$ and $A_j(t')$ and hence it holds for any order of $t$ and $t'$.

\section*{References}
\bibliography{CBrif_OCT_in_AQC}
\bibliographystyle{iopart-num}

\end{document}